\documentclass[aps,prd,reprint,a4paper,showpacs,nofootinbib,superscriptaddress]{revtex4-2}

\usepackage{bbm}
\usepackage{amsmath}
\usepackage{mathtools}

\usepackage{graphicx}
\usepackage{float}
\usepackage{amssymb}

\usepackage{amssymb}
\usepackage{hyperref}
\usepackage{epstopdf}

\usepackage[utf8]{inputenc}
\hypersetup{
    colorlinks=true,
    linkcolor=blue,
    citecolor=red,
}
\usepackage{color}
\usepackage[T1]{fontenc}
\usepackage{txfonts}
\usepackage{mathrsfs}
\usepackage{latexsym}
\usepackage{epsfig}
\usepackage{epstopdf}
\usepackage{epstopdf}
\usepackage{graphicx}
\usepackage{amssymb}
\usepackage{amsmath}
\usepackage{dcolumn}
\usepackage{bm}
\usepackage{color}
\usepackage{comment}
\usepackage{xcolor}
\usepackage{dsfont}
\usepackage{orcidlink}
\usepackage{amsmath}
\usepackage{epsfig}
\usepackage{tabularx}
\usepackage{array}
\usepackage{subfig}
\usepackage{color}
\usepackage{float}
\allowdisplaybreaks
\usepackage{orcidlink}

\newcommand{\Mb}{\bar M}
\newcommand{\gb}{\bar g}
\newcommand{\rb}{\bar r}
\newcommand{\rh}{\bar r_h}
\newcommand{\Gb}{\bar{\mathcal G}}
\newcommand{\G}{\mathcal G}

\usepackage{geometry}
\begin{document}

\title{Central charges as black-hole hair in weakly coupled trace-anomaly gravity}

\author{Francisco Tello-Ortiz \orcidlink{0000-0002-7104-5746}}
\email{francisco.tello@ufrontera.cl}\affiliation{Departamento de Ciencias Físicas, Universidad de La Frontera,
Casilla 54-D, 4811186 Temuco, Chile.}

\author{A. Vásquez\orcidlink{0009-0009-4925-5630}}
\email{avasquez03@ucn.cl}
\affiliation{Unidad de Equipamiento Cient\'ifico (Maini), Universidad Cat\'olica del Norte, Av. Angamos 0610, Antofagasta, Chile.}

\author{Kazuharu Bamba
\orcidlink{0000-0001-9720-8817}}
\email{bamba@sss.fukushima-u.ac.jp}
\affiliation{ Faculty of Symbiotic Systems Science, Fukushima University, Fukushima 960-1296, Japan.}

\author{G. Mustafa \orcidlink{0000-0003-1409-2009}}
\email{gmustafa3828@gmail.com}
\affiliation{Department of Physics, Zhejiang Normal University, Jinhua, Zhejiang 321004, China}

\begin{abstract}
{We study black holes in Gabadadze's weakly coupled effective theory of the gravitational trace anomaly, keeping both the Euler and the Weyl central charges. We prove that a constant anomalyon on any Ricci-flat geometry, the Kerr family included, solves the full theory if and only if the two central charges coincide. Otherwise, the black holes continuously connected to this branch carry scalar hair whose charge is fixed by the surface gravity, the horizon topology and the difference of central charges alone: maximally supersymmetric matter leaves black holes bald, the Standard Model and photon-dominated sectors produce hair of definite sign, and conformal collider bounds confine it to a universal window. Electric charge separates the two anomaly invariants: the hair reverses sign at a critical charge-to-mass ratio set by the ratio of central charges, its extremal value retains only the Weyl charge, and the Euler charge governs the near-horizon growth of the anomalyon. For every unitary matter sector, the hair regular at the event horizon diverges at the Cauchy horizon. Finally, once the higher-derivative Weyl operator is order-reduced, the background, the anomalyon--graviton mixing and all propagation speeds depend only on the difference of central charges, which extends the no-ghost and Laplacian stability conditions to the full anomaly.}
\end{abstract}

\maketitle

\section{Introduction}

The trace anomaly is one of the most robust imprints of quantum fields on curved spacetime~\cite{CapperDuff1974,DeserDuffIsham1976,Duff1977,Duff1994,BirrellDavies}. For a four-dimensional conformal sector the expectation value of the trace of the stress tensor is fixed by two numbers, the Euler (or $a$-type) and Weyl (or $c$-type) central charges, which multiply the Gauss-Bonnet invariant and the square of the Weyl tensor. These coefficients count the light degrees of freedom of the matter sector, they are scheme independent, and they are tightly constrained: $a$ decreases along renormalization-group flows~\cite{KomargodskiSchwimmer2011}, both are positive in unitary theories, and their ratio is confined by conformal collider bounds~\cite{HofmanMaldacena2008}. Their difference $a-c$ is a particularly sharp probe of the spectrum. It vanishes for $\mathcal N=4$ super-Yang-Mills theory, it is negative for scalars and fermions, positive for gauge fields, and in holographic theories it is controlled by higher-curvature corrections in the bulk~\cite{HenningsonSkenderis1998}.

A local diffeomorphism-invariant action reproducing the anomaly was constructed long ago by Riegert and by Fradkin and Tseytlin~\cite{Riegert1984,FradkinTseytlin1984} and later reobtained as a dilaton effective action~\cite{KomargodskiSchwimmer2011} and as a Wess-Zumino term of the conformal group~\cite{GabadadzeTukhashvili2020}. The same structure appears in the regularized four-dimensional Einstein-Gauss-Bonnet theory~\cite{GlavanLin2020,LuPang2020,Kobayashi2020,Fernandes2021,Fernandes2022review}. In all of these realizations the dilaton lacks a quadratic kinetic term and the theory is strongly coupled at arbitrarily low energies~\cite{Bonifacio2020}, a pathology that also shows up as ghost or Laplacian instabilities of the corresponding black holes~\cite{Tsujikawa2022,KaseTsujikawa2023}. Gabadadze showed that the problem is cured by adding a second Einstein-Hilbert term for the Weyl-invariant metric, with a scale $\Mb$ well below the Planck mass, which supplies the dilaton, now called the anomalyon, with a healthy kinetic term without altering the anomaly~\cite{Gabadadze2023}. The theory is weakly coupled below $\Mb$, admits a holographic completion~\cite{GabadadzePorrati2023}, and has been applied to inflation and primordial black holes~\cite{Gabadadze2025} and to Einstein-Gauss-Bonnet-like cosmology~\cite{Tukhashvili2025}; its consistency as an effective field theory has been reviewed in Ref.~\cite{GabadadzeTukhashvili2025}.

Tsujikawa constructed static black holes of this theory sourced by the Euler part of the anomaly and showed that, in contrast with the exact solutions of four-dimensional Einstein-Gauss-Bonnet gravity, {they satisfy the no-ghost and Laplacian stability conditions}~\cite{Tsujikawa2023}. Their scalar hair is of the type sourced by a linear coupling to the Gauss-Bonnet invariant~\cite{SotiriouZhou2014PRL,SotiriouZhou2014PRD,PrabhuStein2018}, a well-studied mechanism for evading the no-hair theorems~\cite{HerdeiroRadu2015}. Those solutions have since been used in phenomenological studies~\cite{Boraho2025}. The Weyl part of the anomaly was switched off in that analysis because it takes the theory outside the Horndeski class~\cite{Horndeski1974,Kobayashi2011}. Since the Weyl charge is never zero in a unitary theory, the black-hole sector of the complete anomaly remained open. {Including it is not a change of coupling: the Weyl term enters the metric equations through fourth-order derivatives of the metric, so a priori it modifies both the background solutions and the structure of their perturbations.}

In this paper we solve it. The central observation is that on Ricci-flat backgrounds the Gauss-Bonnet and Weyl invariants coincide, while the Weyl-tensor contribution to the metric equations is controlled by the Bach tensor, which vanishes on any Einstein space. These two facts turn the full problem into a sharp statement about the central charges{, and they explain why the combination $a-c$ organizes the entire Ricci-flat sector}. We prove that constant-anomalyon Ricci-flat black holes, the Kerr family included, are exact solutions if and only if $a=c$, and that for $a\neq c$ {the black holes continuously connected to this branch carry} anomalyon hair whose charge is fixed by the surface gravity and the difference $a-c$. We derive a covariant master equation for the hair on any background with vanishing Ricci scalar, which gives closed-form results for charged black holes. Electric charge distinguishes the two anomaly invariants and produces a sequence of new effects: a sign inversion of the hair at a critical charge-to-mass ratio fixed by $a/c$, an extremal limit in which the hair measures $c$ alone while $a$ governs a logarithmic growth of the anomalyon at the degenerate horizon, and a positivity property which shows that for every unitary matter sector the hair regular at the event horizon is singular at the Cauchy horizon. {Finally, we show that, once the Weyl operator is order-reduced within the regime of validity of the effective theory, the quadratic action through first order in the anomaly couplings depends on $a-c$ only. Propositions 1 and 2 are exact statements of the full theory, whereas the scalar charge, the charged profiles and the perturbative analysis are obtained at first order in the anomaly couplings, within the regime specified in Sec.~\ref{sec:validity}.}

The paper is organized as follows. Section~\ref{sec:theory} presents the theory with both central charges. Section~\ref{sec:exact} contains the exact results. Section~\ref{sec:vacuum} derives the master equation{, its regime of validity} and the vacuum hair, and connects it with the field content. Section~\ref{sec:charged} treats charged black holes and their interiors. Section~\ref{sec:pert} analyzes linear perturbations, and Sec.~\ref{sec:conc} contains our conclusions. {Explicit derivations for the charged background are collected in Appendix~\ref{app:RN}.} We use the mostly-plus signature and the conventions of Ref.~\cite{Tsujikawa2023}.

\section{Trace-anomaly gravity with both central charges}\label{sec:theory}

For massless fields in an external metric the trace anomaly reads~\cite{Duff1977,BirrellDavies}
\begin{equation}
T_A=-\alpha\,\G+\beta\,W^2,\label{eq:TA}
\end{equation}
with
\begin{align}
\G&=R^2-4R_{\mu\nu}R^{\mu\nu}+R_{\mu\nu\rho\sigma}R^{\mu\nu\rho\sigma},\\
W^2&=\tfrac13R^2-2R_{\mu\nu}R^{\mu\nu}+R_{\mu\nu\rho\sigma}R^{\mu\nu\rho\sigma},
\end{align}
where the scheme-dependent $\Box R$ term has been omitted. The coefficients are the central charges, $\alpha\equiv a$ and $\beta\equiv c$, with the sign conventions of Refs.~\cite{Tsujikawa2023,GabadadzeTukhashvili2025}, in which the anomaly equation in the physical frame reads $2M^2R=a\,\G-c\,W^2$ and both coefficients are positive in unitary theories. For $N_s$ real conformal scalars, $N_f$ Dirac fermions and $N_g$ gauge fields~\cite{Duff1994,GabadadzeTukhashvili2025},
\begin{align}
a&=\frac{N_s+11N_f+62N_g}{360(4\pi)^2},\quad
c=\frac{N_s+6N_f+12N_g}{120(4\pi)^2},\label{eq:ac}\\
a-c&=\frac{-2N_s-7N_f+26N_g}{360(4\pi)^2}.\label{eq:amc}
\end{align}

Introducing the Weyl-invariant metric $\gb_{\mu\nu}=e^{-2\phi}g_{\mu\nu}$, invariant under $g_{\mu\nu}\to e^{2\sigma}g_{\mu\nu}$, $\phi\to\phi+\sigma$, Gabadadze's action is~\cite{Gabadadze2023,Tsujikawa2023}
\begin{align}
S=\int d^4x\sqrt{-\gb}\Big[&\big(M^2e^{2\phi}-\Mb^2\big)\bar R-12M^2e^{2\phi}\bar X\nonumber\\
&-\alpha\big(\phi\Gb-4\bar G^{\mu\nu}\bar\nabla_\mu\phi\bar\nabla_\nu\phi+8\bar X\bar\Box\phi-8\bar X^2\big)\nonumber\\
&+\beta\,\phi\,\bar W^2\Big],\label{eq:action}
\end{align}
with $\bar X=-\tfrac12\gb^{\mu\nu}\partial_\mu\phi\partial_\nu\phi$, $M=M_{\rm Pl}/\sqrt2$ and $\Mb\ll M$. In terms of the physical metric $g_{\mu\nu}$, to which matter couples up to corrections of order $\Mb^2/M^2$~\cite{Gabadadze2023}, the first line becomes $M^2R-\varphi^2R-6(\partial\varphi)^2$ with $\varphi=\Mb e^{-\phi}$: a conformally coupled scalar with a healthy kinetic term. Since $\sqrt{-\gb}\,\bar W^2=\sqrt{-g}\,W^2$, the last term is frame independent.

Varying with respect to $\phi$ at fixed $\gb_{\mu\nu}$ gives the anomaly equation in the physical frame,
\begin{equation}
2M^2R-\alpha\G+\beta W^2=0,\label{eq:scalar}
\end{equation}
and combining it with the trace of the $\gb_{\mu\nu}$ equation gives its counterpart in the Weyl-invariant frame~\cite{Tsujikawa2023},
\begin{equation}
2\Mb^2\bar R-\alpha\Gb+\beta\bar W^2=0.\label{eq:scalarbar}
\end{equation}
The metric equations are
\begin{align}
\big(M^2e^{2\phi}-\Mb^2\big)\bar G_{\mu\nu}=&-2M^2e^{2\phi}\big[\bar\nabla_\mu\phi\bar\nabla_\nu\phi-\bar\nabla_\mu\bar\nabla_\nu\phi\nonumber\\
&+\gb_{\mu\nu}(\bar\Box\phi-\bar X)\big]-\alpha\bar H_{\mu\nu}\nonumber\\
&-2\beta\big(2\bar\nabla^\rho\bar\nabla^\sigma+\bar R^{\rho\sigma}\big)\big(\phi\,\bar C_{\mu\rho\nu\sigma}\big),\label{eq:metric}
\end{align}
where $\bar H_{\mu\nu}$, given in Eq.~(2.17) of Ref.~\cite{Tsujikawa2023}, is built entirely from first and second derivatives of $\phi$. Two properties of Eq.~\eqref{eq:metric} drive everything that follows. First, every term in $\bar H_{\mu\nu}$ contains derivatives of $\phi$. Second, for constant $\phi=\phi_0$ the Weyl contribution reduces to $-4\beta\phi_0\bar B_{\mu\nu}$, with $\bar B_{\mu\nu}=(\bar\nabla^\rho\bar\nabla^\sigma+\tfrac12\bar R^{\rho\sigma})\bar C_{\mu\rho\nu\sigma}$ the Bach tensor, which vanishes on every Einstein space. The field content enters only through the ratio
\begin{equation}
\mathcal A\equiv\frac{a-c}{c}=\frac{a}{c}-1,
\end{equation}
listed in Table~\ref{tab:sectors} for representative sectors. {For any unitary four-dimensional conformal field theory with a unique stress tensor, positivity of the energy flux measured in conformal collider experiments, which follows from the averaged null energy condition~\cite{Faulkner2016,Hartman2017}, implies the bounds}~\cite{HofmanMaldacena2008,Hofman2016} $1/3\le a/c\le 31/18$, that is
\begin{equation}
-\tfrac23\le\mathcal A\le\tfrac{13}{18},\label{eq:HM}
\end{equation}
saturated by free scalars and free Maxwell fields respectively, with narrower windows in supersymmetric theories, $1/2\le a/c\le3/2$ for $\mathcal N=1$ and $1/2\le a/c\le5/4$ for $\mathcal N=2$~\cite{HofmanMaldacena2008,Hofman2016}. {These bounds apply to interacting and free theories alike, the free fields saturating them, and every bound on the scalar charge derived below inherits these assumptions.}

\begin{table}[t]
\caption{Central charges, in units of $1/[360(4\pi)^2]$, for representative conformal sectors, and the sign of the anomalyon charge of a neutral black hole. The Standard Model entry assumes three generations without right-handed neutrinos {and treats all its fields as massless}.}\label{tab:sectors}
\begin{ruledtabular}
\begin{tabular}{lcccc}
Sector & $a$ & $c$ & $a/c$ & ${\rm sign}\,q$\\
\hline
Real scalar & $1$ & $3$ & $1/3$ & $-$\\
Weyl fermion & $11/2$ & $9$ & $11/18$ & $-$\\
Maxwell field & $62$ & $36$ & $31/18$ & $+$\\
$\mathcal N=1$ chiral multiplet & $15/2$ & $15$ & $1/2$ & $-$\\
$\mathcal N=1$ vector multiplet & $135/2$ & $45$ & $3/2$ & $+$\\
$\mathcal N=4$ SYM (per generator) & $90$ & $90$ & $1$ & $0$\\
Standard Model & $1991/2$ & $849$ & $1.173$ & $+$\\
\end{tabular}
\end{ruledtabular}
\end{table}

\section{Exact results}\label{sec:exact}

\emph{Proposition 1.} Let $g_{\mu\nu}$ be a Ricci-flat black-hole metric and $\phi=\phi_0$ constant. Then $(g_{\mu\nu},\phi_0)$ solves the full system \eqref{eq:scalar}--\eqref{eq:metric} if and only if $a=c$.

\emph{Proof.} For constant $\phi$ the metric $\gb_{\mu\nu}=e^{-2\phi_0}g_{\mu\nu}$ is Ricci flat as well. In Eq.~\eqref{eq:metric} the left-hand side vanishes, the bracket and $\bar H_{\mu\nu}$ vanish because they contain derivatives of $\phi$, and the Weyl term is proportional to the Bach tensor of a Ricci-flat metric, which is zero. The metric equations are therefore satisfied identically for every $\alpha$ and $\beta$. On a Ricci-flat metric $R=0$ and $\G=W^2=R_{\mu\nu\rho\sigma}R^{\mu\nu\rho\sigma}$, so Eqs.~\eqref{eq:scalar} and \eqref{eq:scalarbar} both reduce to $(\alpha-\beta)R_{\mu\nu\rho\sigma}R^{\mu\nu\rho\sigma}=0$. For a black hole the Kretschmann scalar does not vanish identically, so the scalar equations hold if and only if $\alpha=\beta$. $\square$

Two consequences follow at once. If $a=c$, the Schwarzschild and Kerr~\cite{Kerr1963} metrics with a constant anomalyon are exact solutions of the complete anomaly-corrected theory, to all orders in the anomaly couplings and with the Weyl term included. {Conversely, if $a\neq c$ the Ricci-flat branch with constant anomalyon is excluded. The black holes continuously connected to it in the anomaly couplings are constructed in Sec.~\ref{sec:vacuum}, where they are shown to carry scalar hair at first order.} This is the content of the no-hair statement of Ref.~\cite{Tsujikawa2023} generalized to the complete anomaly: the relevant coupling is not $a$ but $a-c$.

\emph{Proposition 2.} Let $g_{\mu\nu}$ be an electrovacuum solution with nonzero electromagnetic stress tensor. Then a constant anomalyon is incompatible with Eq.~\eqref{eq:scalar} whenever $a\neq0$.

\emph{Proof.} Maxwell theory is conformally invariant, so it does not enter Eq.~\eqref{eq:scalar}, and its stress tensor is traceless, so $R=0$. Using $\G=W^2-2R_{\mu\nu}R^{\mu\nu}+\tfrac23R^2$, Eq.~\eqref{eq:scalar} becomes $(\alpha-\beta)W^2-2\alpha R_{\mu\nu}R^{\mu\nu}=0$. For $a=c$ the left-hand side is $-2\alpha R_{\mu\nu}R^{\mu\nu}$, which is nonzero wherever the field strength is. $\square$

Hence in a theory with $a=c$ the hair is generated exclusively by electric charge, a point we make quantitative in Sec.~\ref{sec:charged}.

\section{Vacuum hair at leading order}\label{sec:vacuum}

\subsection{Master equation}

We now expand around a background $(g^{(0)}_{\mu\nu},\phi_0)$ that solves the theory with $\alpha=\beta=0$ and has $R^{(0)}=0$. This class contains every Ricci-flat metric and every electrovacuum solution. We write $\phi=\phi_0+\delta\phi$, with $\delta\phi$ and the metric corrections of first order in $\alpha$ and $\beta$.

Using $R=e^{-2\phi}(\bar R-6\bar\Box\phi-6\gb^{\mu\nu}\partial_\mu\phi\partial_\nu\phi)$ at first order, Eqs.~\eqref{eq:scalar} and \eqref{eq:scalarbar} fix both $R^{(1)}$ and $\bar R^{(1)}$ in terms of the background invariants, and their difference is a wave equation for $\delta\phi$. In the physical frame it reads
\begin{equation}
\Box\,\delta\phi=\kappa\,\big(\alpha\,\G-\beta\,W^2\big),\qquad
\kappa\equiv\frac{M^2-\varphi_0^2}{12M^2\varphi_0^2},\label{eq:master}
\end{equation}
with $\varphi_0=\Mb e^{-\phi_0}$ and the right-hand side evaluated on the background. The coefficient $\kappa$ is positive precisely under the no-ghost condition $M^2e^{2\phi_0}>\Mb^2$ of the tensor sector~\cite{Tsujikawa2023}.

The metric correction is sourced by $\delta\phi$ only. Indeed, at first order the $\alpha\bar H_{\mu\nu}$ term in Eq.~\eqref{eq:metric} is proportional to $\alpha\,\delta\phi$, and the Weyl term is proportional either to $\beta$ times the Bach tensor of the first-order metric or to $\beta$ times derivatives of $\delta\phi$; all of these are second order. The linearized metric equations are therefore those of general relativity with a conformally coupled scalar, with $\delta\phi$ as the only source.

Equation~\eqref{eq:master} and this last property imply that on Ricci-flat backgrounds, where $\G=W^2$, the complete first-order solution depends on the anomaly only through $\alpha-\beta$. For Schwarzschild, in the Weyl-invariant frame with $\bar f=-\gb_{tt}$, $\bar h=\gb^{\rb\rb}$, the regular solution is obtained from Ref.~\cite{Tsujikawa2023} by the replacement $\alpha\to\alpha-\beta$:
\begin{align}
\phi&=\phi_0-\frac{(6\rb^2+3\rh\rb+2\rh^2)(M^2-\Mb^2e^{-2\phi_0})}{18M^2\Mb^2\rh\,\rb^3}(\alpha-\beta),\label{eq:phi1}\\
\bar h&=\Big(1-\frac{\rh}{\rb}\Big)\Big[1-\frac{(\rb+2\rh)(\rb+5\rh)}{18\Mb^2\rh\,\rb^3}(\alpha-\beta)\Big],\label{eq:h1}\\
\bar f&=\Big(1-\frac{\rh}{\rb}\Big)\Big[1+\frac{23\rb^2+11\rh\rb+6\rh^2}{18\Mb^2\rh\,\rb^3}(\alpha-\beta)\Big].\label{eq:f1}
\end{align}
We have verified that Eq.~\eqref{eq:phi1} solves Eq.~\eqref{eq:master} transformed to the barred frame. {The amplitude of the vacuum hair is set by $\hat\alpha\equiv|\alpha-\beta|/(\Mb^2\rh^2)$, while the control parameter of the expansion is given in Sec.~\ref{sec:validity}.} Note that the Weyl anomaly does not simply add a new correction to the solution of Ref.~\cite{Tsujikawa2023}: it partially cancels the Euler contribution, and for $a=c$ it removes it entirely, in agreement with Proposition~1.

\subsection{{Validity of the weak-coupling expansion}}\label{sec:validity}

{The first-order results of this paper hold under the following hypotheses, which define the regime of the expansion.}

{\emph{Scale hierarchy.} The theory is weakly coupled below $\Mb$, with $\Mb\ll M$~\cite{Gabadadze2023}. For $\varphi_0\ll M$ one has $\kappa\simeq1/(12\varphi_0^2)=e^{2\phi_0}/(12\Mb^2)$, so the strength of the anomaly source in Eq.~\eqref{eq:master} is set by $\kappa a$ and $\kappa c$.}

{\emph{Curvature.} The expansion is controlled by}
\begin{equation}
{\epsilon\equiv\frac{\kappa\,\max(a,c)}{r_h^2}\ll1,}\label{eq:epsilon}
\end{equation}
{with $r_h$ the horizon radius in the physical frame, which bounds the curvature of the whole exterior region. While $\hat\alpha$ measures the size of the vacuum hair and vanishes for $a=c$, $\epsilon$ bounds each anomaly contribution separately, so the expansion is uniform in the limit $a\to c$.}

{\emph{Field amplitudes.} The anomalyon correction satisfies $|\delta\phi|=\mathcal O(\epsilon)$ and the metric corrections are of relative order $\epsilon$ in every region where the first-order profile is used.}

{\emph{Frequencies and momenta.} In the perturbative analysis of Sec.~\ref{sec:pert}, frequencies and momenta lie well below the cutoff $\Mb$ of the weakly coupled theory, which is the regime in which the Weyl operator is consistently order-reduced.}

{\emph{Horizon regions.} At a nonextremal event horizon all first-order quantities are regular. The expansion ceases to be uniform only in an exponentially thin layer at a degenerate horizon, quantified in Sec.~\ref{sec:charged}, and in a neighborhood of the Cauchy horizon, where the logarithmic growth of Eq.~\eqref{eq:logCauchy} eventually violates the amplitude condition.}

\subsection{Scalar charge}

{We consider a stationary, asymptotically flat black hole with Killing field $\xi^\mu$, a nondegenerate Killing horizon with bifurcation surface $B$, and a stationary anomalyon, $\mathcal L_\xi\delta\phi=0$, that is regular (bounded, with bounded gradient) on $B$ and behaves as $\phi=\phi_\infty-q/r+\mathcal O(r^{-2})$ in the physical frame. Since $\nabla_\mu\xi^\mu=0$ and $\mathcal L_\xi\nabla^\mu\delta\phi=0$, the antisymmetric tensor $F^{\mu\nu}\equiv\xi^\mu\nabla^\nu\delta\phi-\xi^\nu\nabla^\mu\delta\phi$ satisfies $\nabla_\nu F^{\mu\nu}=\xi^\mu\Box\delta\phi$. We integrate this identity over a hypersurface $\Sigma$ extending from $B$ to spatial infinity and apply Stokes' theorem. The contribution of $B$ vanishes because $\xi^\mu=0$ there and $\delta\phi$ is regular, while the sphere at infinity gives $4\pi q$. Using Eq.~\eqref{eq:master}, the scalar charge is}
\begin{equation}
q=\frac{\kappa}{4\pi}\int_\Sigma d^3x\sqrt{-g}\,\big(\alpha\,\G-\beta\,W^2\big).\label{eq:gauss}
\end{equation}
For a static spherically symmetric metric with $-g_{tt}=g^{rr}=f(r)$ the Gauss-Bonnet density is a total derivative,
\begin{equation}
r^2\G=4\,\frac{d}{dr}\big[f'(f-1)\big],\label{eq:GBtotal}
\end{equation}
so that the Euler part of Eq.~\eqref{eq:gauss} depends only on horizon data,
\begin{equation}
\int_{r_+}^{\infty}dr\,r^2\G=4f'(r_+)=8\kappa_H,\label{eq:Eulerpart}
\end{equation}
with $\kappa_H$ the surface gravity. This is the static version of the topological horizon integral of Prabhu and Stein for linear Gauss-Bonnet couplings~\cite{PrabhuStein2018}, who showed that for $\Box\vartheta=-(\alpha/8)\G$ the monopole charge of any stationary, asymptotically flat black hole with compact bifurcation surface $B$ is $\tfrac12\alpha\kappa_H\chi(B)$. On a Ricci-flat background $W^2=\G$, so the Weyl part is topological too, and
\begin{equation}
q=8\kappa\,(a-c)\,\kappa_H\label{eq:qvac}
\end{equation}
for Schwarzschild. {Since the charge of Ref.~\cite{PrabhuStein2018} is obtained from the same Stokes construction, their result is equivalent to}
\begin{equation}
{\frac{1}{4\pi}\int_\Sigma d^3x\sqrt{-g}\,\G=4\kappa_H\,\chi(B)}\label{eq:EulerTop}
\end{equation}
{for every stationary black hole of the class considered above. For a spherical cross section $\chi(B)=2$, and Eq.~\eqref{eq:EulerTop} reduces to Eq.~\eqref{eq:Eulerpart}. Since $W^2=\G$ pointwise on any Ricci-flat metric, the Weyl integral equals the Euler one, and Eq.~\eqref{eq:qvac} holds for the whole Kerr family. With spin parameter $a_{\rm K}$, $r_+=m+\sqrt{m^2-a_{\rm K}^2}$ and $\kappa_H=\sqrt{m^2-a_{\rm K}^2}/(2mr_+)$, it gives}
\begin{equation}
{q_{\rm Kerr}=\frac{4\kappa\,(a-c)\sqrt{m^2-a_{\rm K}^2}}{m\,r_+},}\label{eq:qKerr}
\end{equation}
{which reduces to the Schwarzschild value for $a_{\rm K}=0$.} The scalar charge is therefore universal: it is set by the surface gravity, by the horizon topology and by $a-c$, and it vanishes in the extremal Kerr limit. For Schwarzschild, $\kappa_H=1/(2r_h)$ gives $q=4\kappa(a-c)/r_h$, which reproduces Eq.~(3.22) of Ref.~\cite{Tsujikawa2023} with $\alpha\to\alpha-\beta$ once the change of frame $r_h=e^{\phi_0}\rh$ is taken into account.

\subsection{Field content}

{Equation~\eqref{eq:qvac} makes the scalar charge a direct function of $a-c$.} Several consequences follow from Table~\ref{tab:sectors} and Eq.~\eqref{eq:HM}.

\emph{Supersymmetric bald black holes.} For $\mathcal N=4$ super-Yang-Mills theory, $a=c$ for any gauge group, so Schwarzschild and Kerr with constant anomalyon are exact solutions. More generally, any matter sector with $a=c$, as in large-$N$ theories with an Einstein-gravity holographic dual~\cite{HenningsonSkenderis1998}, leaves vacuum black holes bald.

\emph{Sign of the hair.} Scalars and fermions push $q$ to negative values and gauge fields to positive values, Eq.~\eqref{eq:amc}. Vector-dominated sectors, including the Standard Model, for which $\mathcal A\simeq0.173$, give $q>0$.

\emph{Photon-dominated regime.} For black holes much larger than the Compton wavelengths of all massive species the conformal sector reduces to the photon, which saturates the upper bound in Eq.~\eqref{eq:HM}. Astrophysical black holes therefore carry the maximal positive hair per unit Weyl charge allowed by {the conformal collider bounds}.

\emph{Universal window.} Combining Eqs.~\eqref{eq:HM} and \eqref{eq:qvac}, every {conformal sector obeying the bounds of Eq.~\eqref{eq:HM}} gives
\begin{equation}
-\frac{8}{3}\,\kappa\,c\le q\,r_h\le\frac{26}{9}\,\kappa\,c
\end{equation}
for Schwarzschild black holes. The hair per unit Weyl charge is bounded on both sides, with the extremes attained by free fields.

\section{Charged black holes}\label{sec:charged}

\subsection{Profile and scalar charge}

Consider now the Reissner-Nordstr\"om background in the physical frame, $f=1-2m/r+Q^2/r^2=(1-r_+/r)(1-r_-/r)$, with $r_\pm=m\pm\sqrt{m^2-Q^2}$ and $Q$ the geometric charge parameter. We take the conformal sector to be neutral under the $U(1)$ of the background, as for free fields, so that the anomaly contains no $F_{\mu\nu}F^{\mu\nu}$ term~\cite{GabadadzeTukhashvili2025}. Since Maxwell theory is conformally invariant, the Reissner-Nordstr\"om metric with constant $\phi_0$ solves the theory with $\alpha=\beta=0$, and its curvature invariants are
\begin{equation}
W^2=\frac{48(mr-Q^2)^2}{r^8},\qquad R_{\mu\nu}R^{\mu\nu}=\frac{4Q^4}{r^8},
\end{equation}
with $\G=W^2-2R_{\mu\nu}R^{\mu\nu}$. Equation~\eqref{eq:master} integrates once to
\begin{equation}
\delta\phi'(r)=\frac{\mathcal N(r)}{r^2f(r)},\qquad
\mathcal N(r)=\kappa\int_{r_+}^{r}ds\,s^2\big(\alpha\G-\beta W^2\big),\label{eq:profile}
\end{equation}
where the lower limit is fixed by regularity at the event horizon {(the explicit steps are given in Appendix~\ref{app:RN})}. Both integrals are elementary. Using Eq.~\eqref{eq:GBtotal} and defining
\begin{equation}
\mathcal W(r)\equiv\int_r^\infty ds\,s^2W^2=48\Big[\frac{m^2}{3r^3}-\frac{mQ^2}{2r^4}+\frac{Q^4}{5r^5}\Big],
\end{equation}
one finds
\begin{equation}
\begin{split}
\mathcal N(r)=\kappa\Big\{&4\alpha\big[f'(r)\big(f(r)-1\big)+f'(r_+)\big]\\
&-\beta\big[\mathcal W(r_+)-\mathcal W(r)\big]\Big\}.
\end{split}\label{eq:Nr}
\end{equation}

The scalar charge is $q=\mathcal N(\infty)$. With $y\equiv r_-/r_+\in[0,1]$ it takes the form
\begin{equation}
q=\frac{4\kappa}{r_+}\Big[a\,(1-y)-\frac{c}{5}\big(5-5y+2y^2\big)\Big].\label{eq:qRN}
\end{equation}
The first term is $8\kappa a\kappa_H$, with $\kappa_H=(r_+-r_-)/(2r_+^2)$, in agreement with the topological character of the Euler contribution. The second term is not topological: on a charged background $W^2\neq\G$, and the Weyl contribution depends on the full exterior geometry. Figure~\ref{fig:charge} shows the normalized charge as a function of $Q/m$ for four sectors.

\subsection{Hair inversion}

Writing Eq.~\eqref{eq:qRN} as $q\,r_+/(4\kappa)=(a-c)(1-y)-2cy^2/5$ shows that the charge-induced contribution is negative definite and grows with $y$, while the vacuum contribution decreases. For $a>c$ the hair is positive for small charge and must change sign. It vanishes at
\begin{equation}
y_*=\frac{5}{4}\Big[\sqrt{\mathcal A^2+\tfrac85\mathcal A}-\mathcal A\Big],\qquad
\frac{Q_*}{m}=\frac{2\sqrt{y_*}}{1+y_*}.\label{eq:ystar}
\end{equation}
{This root exists and is unique for $a>c$ (Appendix~\ref{app:RN}).} The critical ratio grows monotonically with $\mathcal A$: it moves to zero as $a\to c^+$ and reaches $Q_*/m\simeq0.986$ for a free Maxwell sector, at the upper end of Eq.~\eqref{eq:HM}; for an $\mathcal N=1$ vector multiplet $Q_*/m\simeq0.978$. For $a\le c$ there is no inversion and $q<0$ for every charge. For $a=c$ the hair is generated exclusively by the electric charge,
\begin{equation}
q\big|_{a=c}=-\frac{8\kappa c}{5r_+}\,y^2,
\end{equation}
in agreement with Proposition~2. The location of the zero of $q$ is a direct readout of the ratio $a/c$.

\subsection{Extremal limit}

\begin{figure}[H]
\centering
\includegraphics[width=\columnwidth]{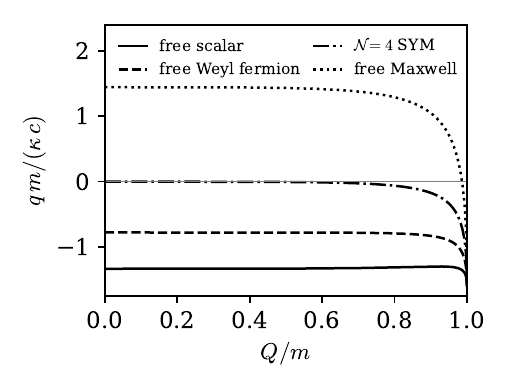}
\caption{Anomalyon scalar charge of a Reissner-Nordstr\"om black hole, normalized by $\kappa c/m$, as a function of the charge-to-mass ratio, for four conformal sectors{: a free real scalar, a free Weyl fermion, $\mathcal N=4$ super-Yang-Mills theory and a free Maxwell field. The curves are first-order results on the fixed Reissner-Nordstr\"om background}. All curves converge to the same value at extremality, where the Euler contribution drops out.}\label{fig:charge}
\end{figure}

At extremality, $y\to1$ and $\kappa_H\to0$, Eq.~\eqref{eq:qRN} gives
\begin{equation}
q\to-\frac{8\kappa c}{5m},\label{eq:qext}
\end{equation}
independently of $a$: the scalar charge of an extremal black hole measures the Weyl central charge alone. {This value is the limit of the asymptotic charge along the nonextremal branch, which is continuous because $q$ depends on the geometry only through Eq.~\eqref{eq:qRN}, smooth in $y$ up to $y=1$.} This is visible in Fig.~\ref{fig:charge}, where all curves meet at $Q=m$.

\begin{figure}[H]
\centering
\includegraphics[width=\columnwidth]{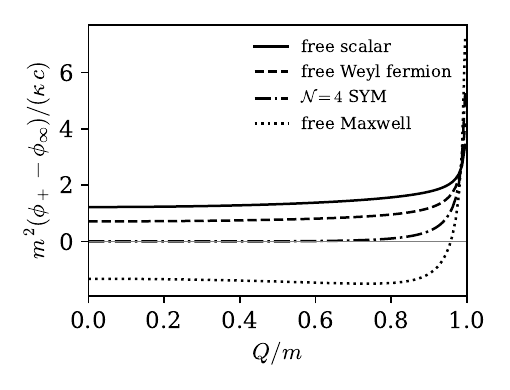}
\caption{Value of the anomalyon at the event horizon relative to its asymptotic value, normalized by $\kappa c/m^2$, as a function of the charge-to-mass ratio. The logarithmic growth near extremality is governed by the Euler charge{, Eq.~\eqref{eq:nearext}; the first-order description holds outside the exponentially narrow window of Eq.~\eqref{eq:extvalid}}.}\label{fig:horizon}
\end{figure}

The Euler charge reappears in the near-horizon behavior. The near-horizon geometry of the extremal solution is $AdS_2\times S^2$ with equal radii, which is conformally flat, so $W^2(r_+)=0$, while $\G(r_+)=-8/m^4$. Since $f$ has a double zero, Eq.~\eqref{eq:profile} gives
\begin{equation}
\delta\phi\simeq-\frac{8\kappa a}{m^2}\,\ln\frac{r-m}{m}\label{eq:extlog}
\end{equation}
near the degenerate horizon. The asymptotic charge is controlled by $c$ and the horizon behavior by $a$, so the two central charges are separately encoded in the extremal hair. Away from extremality the horizon value of the anomalyon is finite and grows logarithmically as $Q\to m$, with a coefficient proportional to $a$, as shown in Fig.~\ref{fig:horizon}.

{The limits $Q\to m$ and $r\to r_+$ do not commute. For fixed $Q<m$ the anomalyon is finite at $r_+$, and near extremality its horizon value behaves as}
\begin{equation}
{\delta\phi(r_+)-\delta\phi(\infty)\simeq-\frac{8\kappa a}{m^2}\ln\frac{2\sqrt{m^2-Q^2}}{m},}\label{eq:nearext}
\end{equation}
{up to a finite term, whereas at $Q=m$ the logarithm of Eq.~\eqref{eq:extlog} diverges at the horizon. The amplitude condition $|\delta\phi|\ll1$ therefore holds at the horizon provided}
\begin{equation}
{1-\frac{Q^2}{m^2}\gg\frac14\exp\Big(-\frac{m^2}{4\kappa a}\Big),}\label{eq:extvalid}
\end{equation}
{and, at extremality, outside the layer $r-m\lesssim m\,e^{-m^2/(8\kappa a)}$. Both regions are exponentially small in the expansion parameter, so the first-order description covers the whole exterior except an exponentially thin layer at the degenerate horizon. The logarithmic near-horizon behavior signals a nonuniform extremal limit and does not by itself establish a globally regular extremal solution: the extremal scalar charge is the formal limit of the nonextremal perturbative branch.}

\subsection{Cauchy horizon}

Regularity at the event horizon uses the only integration constant of Eq.~\eqref{eq:master}, so the behavior at the inner horizon is a prediction. Evaluating Eq.~\eqref{eq:Nr} at $r=r_-$,
\begin{equation}
\mathcal N(r_-)=\kappa\Big\{8a\,(\kappa_++\kappa_-)+c\,\big[\mathcal W(r_-)-\mathcal W(r_+)\big]\Big\},\label{eq:Nminus}
\end{equation}
with $\kappa_\pm=(r_+-r_-)/(2r_\pm^2)$ the surface gravities of the two horizons. Both terms are strictly positive: the first because the surface gravities are positive, the second because $\mathcal W$ is the integral of the nonnegative function $s^2W^2$ and is therefore decreasing. This gives the following result.

\emph{Proposition 3.} For every unitary conformal sector, $a>0$ and $c>0$, the leading-order anomalyon hair of a nonextremal Reissner-Nordstr\"om black hole that is regular at the event horizon diverges at the Cauchy horizon as
\begin{equation}
\delta\phi\simeq-\frac{\mathcal N(r_-)}{2\kappa_-r_-^2}\,\ln\frac{|r-r_-|}{r_+},\qquad \mathcal N(r_-)>0.\label{eq:logCauchy}
\end{equation}

\begin{figure}[H]
\centering
\includegraphics[width=\columnwidth]{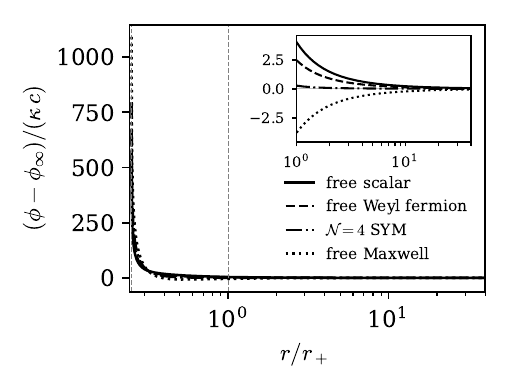}
\caption{Anomalyon profile inside a Reissner-Nordstr\"om black hole with $r_-=r_+/2$, relative to its event-horizon value {and normalized by $\kappa c/r_+^2$}, as a function of the distance to the Cauchy horizon{, for the same four sectors. The curves show the first-order scalar field computed on the fixed Reissner-Nordstr\"om background}. The straight lines on the logarithmic axis display the divergence of Eq.~\eqref{eq:logCauchy}, whose coefficient is positive for every unitary sector. The inset magnifies the exterior region $r\geq r_+$, where the sectors separate according to the sign of $a-c$: the anomalyon approaches its asymptotic value from above for the free scalar and Weyl fermion, from below for the free Maxwell field, and stays close to it for $\mathcal N=4$ super-Yang-Mills theory, whose hair is generated by the electric charge alone.}\label{fig:profile}
\end{figure}

No choice of central charges allowed by unitarity removes the divergence, since a vanishing $\mathcal N(r_-)$ would require $a/c<0$. Figure~\ref{fig:profile} shows the interior profile for the four sectors. {The logarithmic divergence of the anomalyon at the inner horizon establishes a breakdown of the regular perturbative scalar profile on the Reissner-Nordstr\"om background, obtained without introducing dynamical perturbations or choosing a quantum state. Since $\delta\phi'\propto1/|r-r_-|$ enters the anomalyon stress tensor quadratically, it signals that scalar backreaction controls the geometry near $r_-$.} This complements the results of Ref.~\cite{Arrechea2024}, where states of the anomaly-induced effective action were found to be incompatible with regularity at both horizons, the recent argument that the trace anomaly enforces strong cosmic censorship through the stress tensor of the conformal fields~\cite{Easson2025}, and the no-Cauchy-horizon theorems established for charged scalar hair~\cite{Hartnoll2020,Cai2021}, and it adds a static, state-independent {perturbative obstruction} to the classical instability of inner horizons~\cite{PoissonIsrael1990}.

\section{Linear perturbations}\label{sec:pert}

The Weyl term takes the theory outside the Horndeski class, so the stability criteria of Refs.~\cite{KMS2012,KMS2014,KaseTsujikawa2022} cannot be applied directly. We show instead that, to first order in the anomaly couplings {and after order reduction of the Weyl operator}, the quadratic action reduces to a Horndeski one in which $\alpha$ and $\beta$ enter through $\alpha-\beta$ only. {This establishes the absence of ghosts and the Laplacian stability of the perturbations, the two properties controlled by the coefficients of the quadratic action.}

Expand the action \eqref{eq:action} to second order in $h_{\mu\nu}=\delta\gb_{\mu\nu}$ and $\pi=\delta\phi$ around the first-order vacuum solution of Sec.~\ref{sec:vacuum}. Through first order in the couplings there are three contributions{: the Einstein and conformal-scalar sector, the Euler and Weyl mixings, and the Weyl operator}.

(i) The $\alpha=\beta=0$ part of the action evaluated on the corrected background. By Sec.~\ref{sec:vacuum} the background depends only on $\alpha-\beta$.

(ii) The explicit anomaly terms evaluated on the zeroth-order background, which is Ricci flat with constant $\phi_0$. In the Euler sector, $\phi_0\Gb$ is a total derivative to all orders, $\bar G^{\mu\nu}$ vanishes and the remaining terms are cubic or quartic in $\pi$, so the only quadratic contribution is the mixing $-\alpha\,\pi\,\delta_1(\sqrt{-\gb}\,\Gb)$. In the Weyl sector, $\delta_1(\sqrt{-\gb}\,\bar W^2)=\delta_1(\sqrt{-\gb}\,\Gb)$ on a Ricci-flat background, because the difference is quadratic in the Ricci tensor. The two mixings combine into
\begin{equation}
-(\alpha-\beta)\,\pi\,\delta_1\big(\sqrt{-\gb}\,\Gb\big),\label{eq:mixing}
\end{equation}
where $\delta_1(\sqrt{-\gb}\Gb)$ is linear in $h_{\mu\nu}$ and proportional to $\bar R^{\mu\alpha\nu\beta}\bar\nabla_\alpha\bar\nabla_\beta h_{\mu\nu}$ up to total derivatives. {The discarded boundary terms come from the total-derivative character of $\sqrt{-\gb}\,\Gb$ and from integrations by parts, and they vanish for perturbations that are regular on the future horizon and decay at spatial infinity, the boundary conditions of the stability and quasinormal-mode problems. The remaining Euler terms of Eq.~\eqref{eq:action}, quadratic or higher in $\partial\phi$, contribute to the quadratic action only through the background gradient or the first-order Einstein tensor, and are therefore of second order.}

(iii) {The constant-coefficient operator $\beta\phi_0\,\delta_2(\sqrt{-g}\,W^2)$. Since $\sqrt{-g}\,\G$ is a total derivative and $W^2=\G+2R_{\mu\nu}R^{\mu\nu}-\tfrac23R^2$, on a Ricci-flat background}
\begin{equation}
{\delta_2\big(\sqrt{-g}\,W^2\big)=2\sqrt{-g}\Big[R^{(1)}_{\mu\nu}R^{(1)\mu\nu}-\tfrac13\big(R^{(1)}\big)^2\Big]+\text{total derivatives}}\label{eq:Weylop}
\end{equation}
{holds exactly, because every other term carries a power of the background Ricci tensor. At zeroth order in the couplings, the linearized equations of general relativity with the conformal scalar $\varphi$ give $R^{(1)}=0$, hence $\Box\delta\varphi=0$, and}
\begin{equation}
{R^{(1)}_{\mu\nu}\propto\frac{\varphi_0}{M^2-\varphi_0^2}\,\nabla_\mu\nabla_\nu\delta\varphi+E_{\mu\nu},}\label{eq:R1}
\end{equation}
{where $E_{\mu\nu}$ is a combination of the linearized equations of motion that vanishes on shell; for tensor modes $R^{(1)}_{\mu\nu}=E_{\mu\nu}$. On a Ricci-flat background, $\int\sqrt{-g}\,\nabla_\mu\nabla_\nu\delta\varphi\,\nabla^\mu\nabla^\nu\delta\varphi=\int\sqrt{-g}\,(\Box\delta\varphi)^2$ up to boundary terms, because commuting the covariant derivatives produces only $R_{\mu\nu}\nabla^\mu\delta\varphi\nabla^\nu\delta\varphi$. Every term of Eq.~\eqref{eq:Weylop} is then quadratic in the leading-order equations of motion and is removed at this order by a field redefinition~\cite{Endlich2017,CanoRuiperez2019}. The operator therefore introduces no new propagating branch for frequencies and momenta below the cutoff of Sec.~\ref{sec:validity}, and since the argument does not involve $a-c$, it holds uniformly as $a\to c$.}

{It follows that, after this order reduction, every first-order quantity of the linear theory depends on the anomaly through $\alpha-\beta$ only, uniformly in the limit $a\to c$.} In particular, the odd-sector coefficients $\bar{\mathcal G}$, $\bar{\mathcal F}$ and $\bar{\mathcal H}$ of Ref.~\cite{Tsujikawa2023} receive first-order corrections only through the background, hence through $\alpha-\beta$, and their ratios, the odd-sector speeds, are corrected at second order. The angular propagation speeds of the even sector, the only ones corrected at first order~\cite{Tsujikawa2023}, become
\begin{equation}
\bar c^{\,2}_{\pm,{\rm even}}=1\pm\frac{2\sqrt3\,\rh\,e^{-\phi_0}}{M\Mb\,\rb^3}\,(\alpha-\beta)+\mathcal O(\epsilon^2),\label{eq:cpm}
\end{equation}
with $\epsilon$ the order of the anomaly couplings. {The no-ghost and Laplacian stability conditions therefore hold within the regime of Sec.~\ref{sec:validity} for any sign of $a-c$, so the conditions established in Ref.~\cite{Tsujikawa2023} for $\beta=0$ extend to the complete anomaly.}

{For $a=c$ the mixing \eqref{eq:mixing} vanishes and Ricci-flat black holes with a constant anomalyon remain exact solutions. Within the low-energy regime where the higher-derivative Weyl operator is consistently order-reduced, their leading linear response, including the quasinormal spectrum, agrees with that of general relativity coupled to a conformal scalar on Schwarzschild, which satisfies the no-ghost and Laplacian conditions~\cite{Tsujikawa2023}, up to corrections of order $\epsilon^2$.}

The even-sector kinetic coefficient of the Horndeski part of the action admits a transparent form. The second-order result of Ref.~\cite{Tsujikawa2023} can be rewritten exactly as
\begin{equation}
\bar{\mathcal K}=\frac{6M^2\Mb^2\,\rb^2\phi'^2}{M^2-\Mb^2e^{-2\phi_0}},\label{eq:K}
\end{equation}
with $\phi'$ the leading-order gradient \eqref{eq:phi1}. The positivity of $\bar{\mathcal K}$ is therefore the no-ghost condition of the conformal scalar of $S_{R\bar R}$ multiplied by the square of the hair, and the Horndeski part evaluated on our background gives $\bar{\mathcal K}\propto(\alpha-\beta)^2$. The appearance of the anomaly coupling in $\bar{\mathcal K}$ reflects the gauge used in Refs.~\cite{KMS2014,KaseTsujikawa2022}, which fixes $\delta\phi=0$ and becomes singular when the background gradient vanishes, rather than a loss of the anomalyon kinetic term: for $a=c$ the anomalyon propagates with the healthy kinetic term supplied by $S_{R\bar R}$.

\section{Conclusions}\label{sec:conc}

We have generalized the black-hole sector of the weakly coupled trace-anomaly framework to incorporate the full trace anomaly, including contributions from both the Euler and Weyl central charges.The guiding principle is that on Ricci-flat backgrounds the two anomaly invariants coincide while the Weyl contribution to the metric equations is governed by the Bach tensor. This reformulates the problem as a statement about the field content: {Ricci-flat black holes with a constant anomalyon, including the Kerr metric, are exact solutions to all orders if and only if the central charges coincide; otherwise this branch is excluded and the black holes continuously connected to it carry nontrivial hair.}

The hair of vacuum black holes is universal. It depends on the surface gravity, on the horizon topology and on the difference of central charges, and nothing else. {The scalar charge thereby encodes the light spectrum of the conformal sector:} maximally supersymmetric matter leaves them bald, scalars and fermions produce hair of one sign and gauge fields of the other, the Standard Model and the photon-dominated regime of astrophysical black holes give positive hair, and {the conformal collider bounds confine} the hair per unit Weyl charge to a finite window whose edges are free fields.

Electric charge separates the two central charges. The hair changes sign at a critical charge-to-mass ratio that measures their ratio, the extremal hair retains only the Weyl charge while the Euler charge fixes the logarithmic behavior at the degenerate horizon, and in theories with equal central charges the hair is purely electric in origin. Inside the black hole, the positivity of both central charges guarantees that hair regular at the event horizon is singular at the Cauchy horizon, a static and state-independent {perturbative} obstruction that the anomaly imposes on the inner horizon. {Near extremality, the first-order description remains valid everywhere outside an exponentially thin layer at the degenerate horizon.}

At the level of linear perturbation theory, {once the higher-derivative Weyl operator is order-reduced,} the full anomaly coincides, to first order, with its Euler contribution, provided that the coupling constant is replaced by the difference between the corresponding central charges. {This extends the no-ghost and Laplacian stability conditions of the hairy solutions to the full anomaly, uniformly in the limit of equal central charges, where the leading linear response of vacuum black holes coincides with that of general relativity with a conformal scalar.}

Several directions follow naturally. A backreacted or nonperturbative treatment of the inner region will determine the ultimate fate of the Cauchy horizon, and a full modal analysis, including the quasinormal spectrum at second order, will complete the stability picture. The second-order solution, where the Weyl term contributes through the Bach tensor of the first-order metric and the Euler and Weyl charges first enter independently in the vacuum sector, can be obtained with order reduction and would give the leading effect that distinguishes $a$ from $c$ without electric charge. The scalar dipole radiation of binaries, which depends on the charge-to-mass ratios and thus on $a-c$, the inclusion of conformal sectors charged under the background gauge field, where the anomaly acquires an additional $F_{\mu\nu}F^{\mu\nu}$ term controlled by the beta function, and the role of a possible anomalyon mass at large distances~\cite{GabadadzeTukhashvili2025} are the next steps toward {connecting the scalar charge with concrete observables and thus toward} observational tests of the central charges with black holes.

\section*{Acknowledgments}
The work of Kazuharu Bamba was supported in part by the JSPS KAKENHI Grant No.~25KF0176 and a grant-in-aid of academic research of the Yamaguchi Scholarship Foundation. 

\appendix

\section{{Explicit derivations for the Reissner-Nordstr\"om background}}\label{app:RN}

This appendix collects the explicit steps behind the results of Sec.~\ref{sec:charged} for the Reissner-Nordstr\"om background. We first list the curvature invariants that source the anomalyon and verify the total-derivative form of the Gauss-Bonnet density, Eq.~\eqref{eq:GBtotal}. We then integrate the master equation~\eqref{eq:master}, fix the integration constant by regularity at the event horizon and extract the scalar charge, Eq.~\eqref{eq:qRN}. Finally, we prove the existence and uniqueness of the sign inversion for $a>c$, treat the cases $a=c$ and $a<c$ separately, and derive the behavior of the anomalyon at the inner horizon, Eqs.~\eqref{eq:Nminus} and \eqref{eq:logCauchy}.

{\emph{Curvature invariants.} For $f=1-2m/r+Q^2/r^2$ the only nonvanishing Weyl scalar is $\Psi_2=-(mr-Q^2)/r^4$, so $W^2=48\Psi_2^2=48(mr-Q^2)^2/r^8$. The Maxwell stress tensor gives $R^t{}_t=R^r{}_r=-Q^2/r^4$ and $R^\theta{}_\theta=Q^2/r^4$, with the remaining angular component equal to $R^\theta{}_\theta$. Hence $R=0$ and $R_{\mu\nu}R^{\mu\nu}=4Q^4/r^8$. For $R=0$, $\G=W^2-2R_{\mu\nu}R^{\mu\nu}$, which gives}
\begin{equation}
{\G=\frac{8\big(6m^2r^2-12mQ^2r+5Q^4\big)}{r^8}.}
\end{equation}
{With $f-1=-2m/r+Q^2/r^2$ and $f'=2m/r^2-2Q^2/r^3$, a direct computation gives $4[f'(f-1)]'=48m^2/r^4-96mQ^2/r^5+40Q^4/r^6=r^2\G$, which is Eq.~\eqref{eq:GBtotal}.}

{\emph{Integration and regularity.} For a static field $\Box\delta\phi=r^{-2}(r^2f\delta\phi')'$, so Eq.~\eqref{eq:master} integrates to $r^2f\delta\phi'=\mathcal N(r)+C$. Near the event horizon $f\simeq2\kappa_+(r-r_+)$ and $\mathcal N(r)=\mathcal O(r-r_+)$, so a bounded gradient at $r_+$ requires $C=0$; any $C\neq0$ produces $\delta\phi\simeq C\ln(r-r_+)/(2\kappa_+r_+^2)$. For $r\to\infty$, $f\to1$ and $r^2\delta\phi'\to\mathcal N(\infty)$, so $\delta\phi=\phi_\infty-\mathcal N(\infty)/r+\mathcal O(r^{-2})$ and $q=\mathcal N(\infty)$.}

{\emph{Scalar charge.} The Euler part gives $4\alpha[f'(f-1)]_{r_+}^{\infty}=4\alpha f'(r_+)$, since $f(r_+)=0$ and $f'(f-1)\to0$ at infinity. With $m=r_+(1+y)/2$ and $Q^2=r_+^2y$, one has $f'(r_+)=(1-y)/r_+$ and $\mathcal W(r_+)=4(5-5y+2y^2)/(5r_+)$, which yields Eq.~\eqref{eq:qRN}.}

{\emph{Sign inversion.} Define $g(y)\equiv q\,r_+/(4\kappa)=(a-c)(1-y)-\tfrac25cy^2$ for $y\in[0,1]$. For $a>c$, $g(0)=a-c>0$, $g(1)=-2c/5<0$ and $g'(y)=-(a-c)-\tfrac45cy<0$, so $g$ is strictly decreasing and has exactly one root, the positive solution of $y^2+\tfrac52\mathcal Ay-\tfrac52\mathcal A=0$, which is Eq.~\eqref{eq:ystar}. For $a=c$, $g(y)=-\tfrac25cy^2$, which vanishes only for the neutral black hole. For $a<c$, both terms of $g$ are nonpositive and never vanish simultaneously, so $g<0$ for every $y\in[0,1]$. Finally, $dy_*/d\mathcal A=\tfrac54\big[(\mathcal A+\tfrac45)/\sqrt{\mathcal A^2+\tfrac85\mathcal A}-1\big]>0$, and $Q_*/m=2\sqrt{y_*}/(1+y_*)$ increases with $y_*\in(0,1)$, so the critical ratio grows monotonically with $\mathcal A$.}

{\emph{Inner horizon.} Since $f'(r_\pm)=\pm2\kappa_\pm$ and $f(r_-)=0$, the Euler part of Eq.~\eqref{eq:Nr} at $r_-$ is $4\alpha[f'(r_+)-f'(r_-)]=8\alpha(\kappa_++\kappa_-)$, and the Weyl part is $\beta[\mathcal W(r_-)-\mathcal W(r_+)]$, positive because $\mathcal W$ is decreasing. This is Eq.~\eqref{eq:Nminus}. Near $r_-$, $r^2f\simeq-2\kappa_-r_-^2(r-r_-)$, and integrating $\delta\phi'=\mathcal N/(r^2f)$ gives Eq.~\eqref{eq:logCauchy}.}

\end{document}